\documentclass[conference]{IEEEtran}
\IEEEoverridecommandlockouts

\usepackage{multirow}
\usepackage{booktabs}
\usepackage{bbm}
\usepackage{tabularx}
\usepackage{hyperref}
\usepackage{makecell}

\usepackage{bbding}
\usepackage{pifont}

\usepackage{cite}
\usepackage{amsmath,amssymb,amsfonts}
\usepackage{algorithmic}
\usepackage{graphicx}
\usepackage{textcomp}
\usepackage{xcolor}
\def\BibTeX{{\rm B\kern-.05em{\sc i\kern-.025em b}\kern-.08em
    T\kern-.1667em\lower.7ex\hbox{E}\kern-.125emX}}
\begin{document}

\title{Codebook Configuration for RIS-aided Systems via Implicit Neural Representations\\
 }


\author{
\IEEEauthorblockN{Huiying Yang, Rujing Xiong, Yao Xiao, Zhijie Fan, Tiebin Mi, Robert Caiming Qiu, Zenan Ling\textsuperscript{\dag}}
\IEEEauthorblockA{
EIC, Huazhong University of Science and Technology, Wuhan 430074, China.\\ 
Ezhou Industrial Technology Research Institute, Huazhong University of Science and Technology, Wuhan 430074, China. \\ 
\textsuperscript{\dag} Corresponding Author. Email: lingzenan@hust.edu.cn}
\thanks{This work was supported by the National Natural Science Foundation of China (via fund NSFC-12141107), the Key Research and Development Program of Hubei (2021BAA037) and of Guangxi (GuiKe-AB21196034). The datasets and code are available at: \url{https://github.com/HUSTGSNeRF/Codebook_Inr}.}
}
\maketitle

\begin{abstract}

Reconfigurable Intelligent Surface (RIS) is envisioned to be an enabling technique in 6G wireless communications. By configuring the reflection beamforming codebook, RIS focuses signals on target receivers to enhance signal strength. In this paper, we investigate the codebook configuration for RIS-aided communication systems. 
We formulate an implicit relationship between user's coordinates information and the codebook from the perspective of signal radiation mechanisms, and introduce a novel learning-based method, implicit neural representations (INRs), to solve this implicit coordinates-to-codebook mapping problem. Our approach requires only user's coordinates, avoiding reliance on channel models.
Additionally, given the significant practical applications of the 1-bit RIS, we formulate the 1-bit codebook configuration as a multi-label classification problem, and propose an encoding strategy for 1-bit RIS to reduce the codebook dimension, thereby improving learning efficiency. Experimental results from simulations and measured data demonstrate significant advantages of our method.


\end{abstract}


\section{Introduction}

Reconfigurable Intelligent Surfaces (RISs) have recently emerged as a promising technology for expanding the coverage of wireless networks and enhancing transmission quality~\cite{wu2019towards,di2020smart}. Generally, the RIS is a type of programmable metasurface composed of a mass of passive elements. Each element can independently adjust the phase of incident electromagnetic waves in order to redirect signals to target receivers. The matrix, which consists of the phase shifts of each element, is called the codebook. Adjusting the configuration of codebook to unlock the full potential of the RIS in signal strength enhancement is of great significance.


Extensive research works have focused on studying the configuration of the codebook. On the one hand, many previous work are model-based. Through the quantitative analysis of the interaction between RIS systems and wireless channel environments, the problem is reformulated as a non-convex optimization problem which takes the codebook as the decision variable and is solved by an optimization algorithm, such as integer programming~\cite{zhang2022configuringAPX}, semidefinite relaxation~\cite{wu2019intelligentSDR}, and manifold optimization~\cite{xiong2023risMAN}. 
However, model-based methods encounter practical challenges owing to the assumption and simplification of channel models, the complexities of optimization and the necessity of extra information of source device. On the other hand, learning-based methods acquire codebook configuration from collected data. \cite{abdallah2023deep} employs reinforcement learning algorithms to configure the codebook based on feedback of signal strength from receivers. Nevertheless, these approaches require real-time feedback, which is difficult to attain in practice. Furthermore, ~\cite{huang2019indoor} proposes a multilayer perceptron (MLP) that utilizes simulation data to learn a mapping from 2D coordinates to the codebook. However, due to the limited expressive power of MLP, the performance is inadequate to meet the demands in practical scenarios.


In this paper, we identify an implicit relationship between user position and the codebook from the perspective of signal radiation mechanisms. We introduce a novel learning-based method, \textit{implicit neural representations} (INRs)~\cite{sitzmann2020implicit,FourierMapping}, to solve this implicit coordinates-to-codebook mapping problem. 
Using user's 3D coordinates information, the proposed model is trained on collected data to learn the corresponding codebook. During inference, the learned model can accurately configure the codebook using only user's 3D coordinates. This method avoids reliance on specific channel models, requires no additional source information, and involves relatively low time overhead, making it easily deployable in practice.

Explicitly obtaining the optimal codebook configuration is challenging due to the complex characteristics of electromagnetic waves~\cite{mi2023towards}.
In this context, INR appears particularly suited for this problem. 
INR enables learning the implicit mapping between the coordinates and target signals. 
It has demonstrated remarkable benefits in tasks involving implicit signal modeling, such as image restoration~\cite{tucker2020single} and  Novel View Synthesis (NVS)~\cite{mildenhall2021nerf}, which involves modeling light field signals. 
Similarly, we employ the INR technique for modeling electromagnetic wave signals, as both are inherently governed by wave equations~\cite{orekondy2022winert}.
For the first time, we introduce the INR method to the task of modeling the codebook configuration of RIS. This method demonstrates its significant advantages in optimizing codebook configuration.

Moreover, we investigate the 1-bit RIS codebook configuration problem, 
a matter of significant importance in practical scenarios~\cite{xiong2022optimal,pei2021ris}. Building upon the implicit representation of codebook, we formulate the problem as a multi-label classification task for the binarization of codebook elements. Additionally,  we develop an encoding method that substantially  reduce the dimension of the 1-bit RIS codebook, thereby increasing training efficiency and  mitigating overfitting.

Our contributions are  summarized as follows.
\begin{itemize}
\item We identify an implicit relationship between user’s coordinates and the codebook from the perspective of signal radiation mechanisms, and propose a novel INR-based method to solve this implicit mapping problem.
\item Given the significance of practical 1-bit RIS, we formulate the 1-bit codebook configuration as a multi-label classification problem, and propose an encoding strategy for 1-bit RIS to reduce the codebook dimension.
\item Our experiments involve the utilization of both simulated and measured data. The results illustrate the notable advantages of proposed the method.
\end{itemize}


\section{Methodology} \label{sec:PROBLEM ANALYSIS AND SYSTEM MODEL}
\begin{figure}
  \centering
  \includegraphics[width = 0.44\textwidth, trim=180 130 205 100, clip]{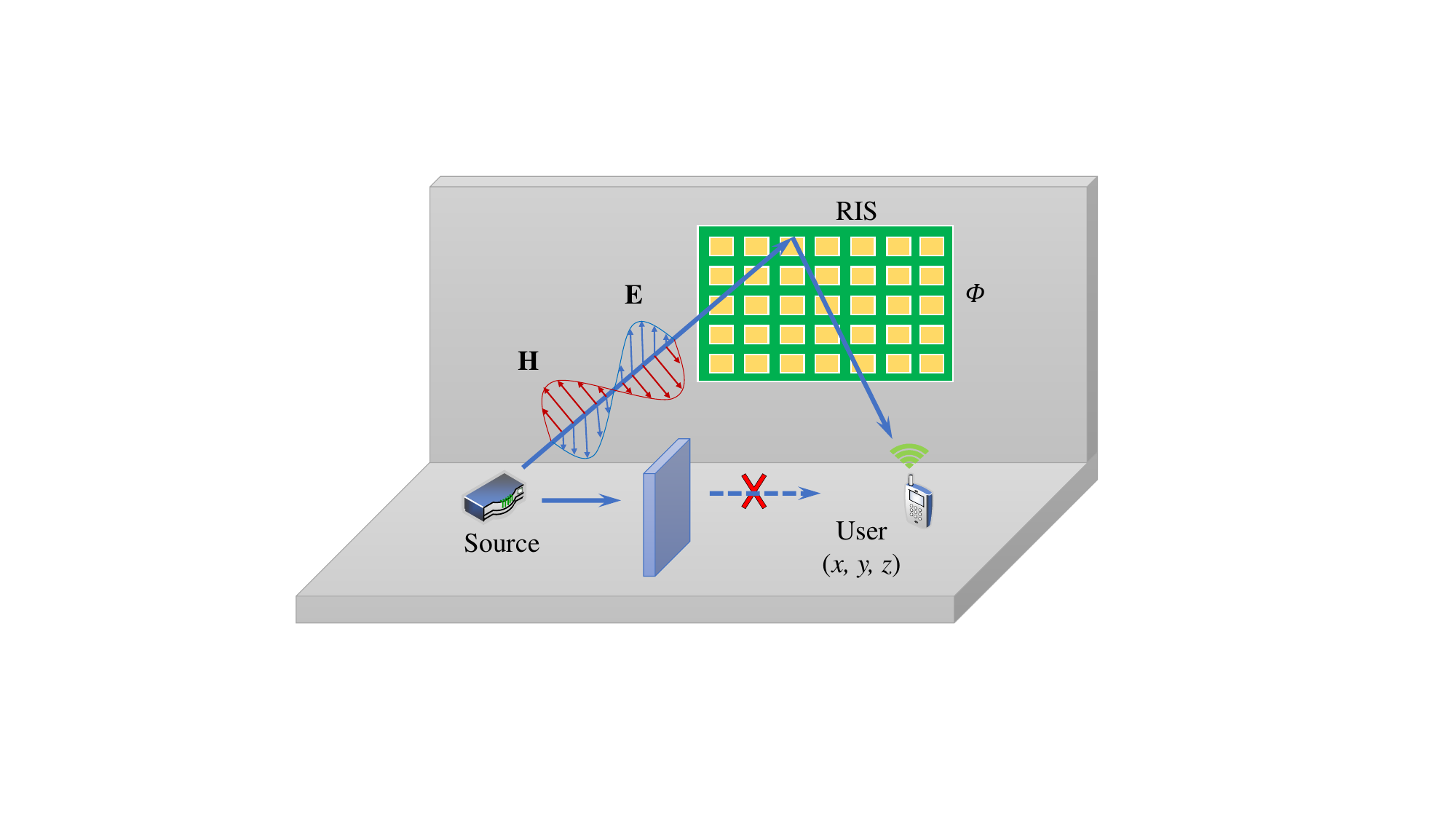}
  \caption{A practical scenario of RIS-aided system. Our objective is to configure the optimal codebook solely based on the user coordinates.}
\label{fig:Practical Scenarios and Optimal Codebook.pdf}
\vspace{-0.2cm}
\end{figure}

In this paper, we explore a practical scenario involving a RIS-assisted communication system, as depicted in Fig.~\ref{fig:Practical Scenarios and Optimal Codebook.pdf}. We consider a fixed spatial configuration for the signal source and the RIS, while allowing for the user's mobility. In this specific scenario, user positioned behind obstacles encounters insufficient signal strength via the line-of-sight link. The RIS comprises \(M\) rows and \(N\) columns of elements, collectively forming a matrix of size \(M \times N\) known as the codebook. The objective is to determine the optimal codebook to maximize the user's signal strength at any position.
\subsection{Problem Formulation}

We approach this problem from the foundational perspective of electromagnetic radiation mechanisms. The electromagnetic wave signal generated at the source propagates in space as a continuous and differentiable function. 
Each element in the codebook performs phase modulation on the incident electromagnetic wave to enhance the signal strength at the target position. This indicates that there is an implicit relationship between the codebook and the coordinates.
we formulate the implicit model to learn this mapping relationship, denoted as

\begin{align}
F\left(\mathbf{v}, \Phi, \nabla_{\mathbf{v}} \Phi, \nabla_{\mathbf{v}}^{2} \Phi, \ldots\right) & = 0, \quad \Phi: \mathbf{v} \mapsto \Phi(\mathbf{v})
\label{eq:formulation}
\end{align}
where $\mathbf{v}:=(x,y,z)$ indicates the user's coordinates and $\Phi$ denotes the RIS codebook configuration.
This implicit problem formulation takes user coordinates, $\mathbf{v}$, as input and, 
the derivatives of $\Phi$ represent the spatial derivatives of signals corresponding with these user coordinates. Our goal is then to learn a neural
network that parameterizes $\Phi$ to map $\mathbf{v}$  to certain quantity of interest while satisfying the constraint presented in~\eqref{eq:formulation}. In this way, the relationship between the coordinates $\mathbf{v}$ and the codebook $\Phi$ is implicitly defined by $F$. 

Due to the complex characteristics of electromagnetic waves, the  implicit mapping defined in~\eqref{eq:formulation} is always complicated and hard to resolve.
The previous methods involve optimizing with signal strength as the objective, constructing a non-convex optimization problem with multiple interdependent parameters, and designing optimization algorithms for resolution.
However, these methods heavily rely on the accurate construction of channel models, which often involve assumptions and simplifications that may not yield satisfactory results in practice.
In this paper, we avoid assuming complex channel models and instead, directly use the real coordinates-to-codebook mapping data to efficiently learn~\eqref{eq:formulation}. In the end, the learned model can accurately configure the codebook of RIS using only user's 3D coordinates.

\subsection{Implicit Neural Representation}
\begin{figure*}
  \centering
  \includegraphics[width = 0.90\textwidth,trim=30 140 95 120, clip]{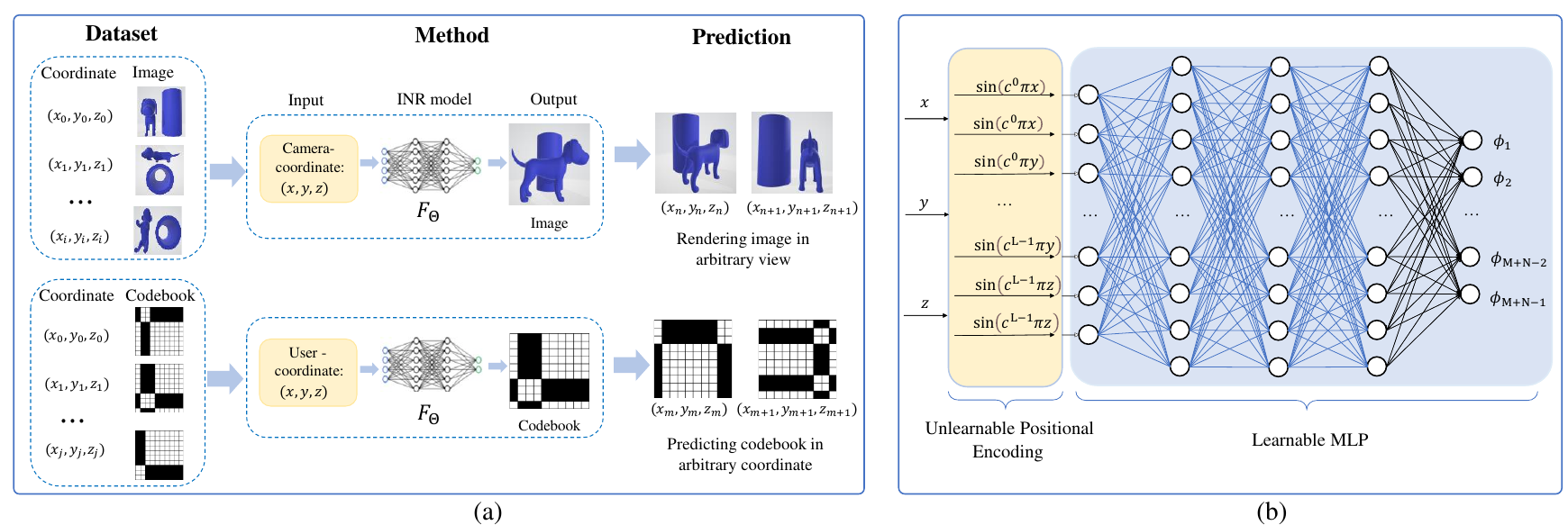}
  \caption{(a) The similarity between the NVS and the codebook configuration. (b) The architecture of INR. It includes six layers: the input layer, four hidden layers, and the output layer, with dimensions of $6L$, 128, 128, 128, 128 and $M+N-1$, respectively. }
\label{fig:similarity comparison}
\vspace{-0.2cm}
\end{figure*}


Based on~\eqref{eq:formulation}, 
our goal is to learn the implicit relationship between coordinates and the codebook.
INR has recently emerged as a potential tool for tasks involving the implicit mapping of coordinates to desired outputs, such as the NVS task~\cite{mildenhall2021nerf}. In Fig.~\ref{fig:similarity comparison}(a), we note the similarity between this problem and the NVS task. 
Additionally, we recognize that the Fourier feature mapping embedded in the INR model structure correlates to the form of the electric signals to a certain extent\cite{FourierMapping}. Motivated by these parallels, we propose employing INR to learn the mapping representation between the user's coordinates and the corresponding codebook. 
Specifically, INR model consists of two primary parts: an unlearnable positional encoding (PE) and a learnable MLP, as illustrated in Fig.~\ref{fig:similarity comparison}(b).
We characterize the codebook-solving model as
\begin{equation}
\Phi=\Psi \circ \Gamma(\mathbf{v})
\vspace{-0.08cm}
\end{equation}
where $\Gamma(\mathbf{v})$ denotes the PE and $\Psi$ denotes a learnable MLP.

\subsubsection{Positional Encoding} The PE mapping function $\Gamma(\cdot)$, is utilized to enhance the INR's capacity to accommodate complex components of signals according to the Fourier transform. Specifically, the PE takes the normalized coordinates $\mathbf{v}:=(x, y, z)$ as input and executes upscaling operation as
\begin{equation}
\Gamma(\mathbf{v})=\left(\begin{array}{l}
\sin (\pi \mathbf{C} \cdot \mathbf{v}) \\
\cos (\pi \mathbf{C} \cdot \mathbf{v}) 
\end{array}\right)
\vspace{-0.08cm}
\end{equation}
where $\mathbf{C}$ represents a Fourier mapping matrix defined as $\mathbf{C}=\left(c^{0}, c^{1}, \cdots, c^{L-1}\right)$, where $c$ denotes the basic frequency, and $L$ is the length of the mapping space. 

As a result, the input dimension of the MLP is expanded from 3 to 6$L$. This expansion can be interpreted as a variation of the Fourier series, aligning with the characteristics of electromagnetic signals~\cite{FourierMapping}. Therefore, INR employs positional encoding to grasp the propagation characteristics of the electric field, thereby enhancing the learning capability.

\subsubsection{Multi-Layer Perceptron}
The INR employs a MLP neural network structure. MLP serves as a fundamental component of deep learning and finds extensive application in tasks such as classification and prediction. The process of forward propagation can be expressed as
\begin{equation*}
\Psi(\mathbf{x})=\sigma\left(\mathbf{W}^{(l)} \cdots\left(\sigma\left(\mathbf{W}^{(1)} \cdot \mathbf{x}+\mathbf{b}^{(1)}\right)+\cdots \mathbf{b}^{(l)}\right)\right)
\label{eq:mlp}
\vspace{-0.07cm}
\end{equation*}

where $\sigma(\cdot)$ represents the activation function, $l$ is the number of layers, and $\mathbf{W}^{(j)}$ and $\mathbf{b}^{(j)}$, where $j=1,2, \cdots, l$,
represent the network weights. Based on this, the architecture intended for learning the mapping relationship from coordinates to the codebook is depicted in Fig.~\ref{fig:similarity comparison}(b).

\subsection{$1$-bit Codebook Configuration}
\subsubsection{ Multi-Label Classification}
 
Theoretically, the elements within the codebook are deemed continuous, indicating their ability to continually adjust the phase of incident electromagnetic waves. However, in practical applications, the characteristics of electromagnetic waves undergo discrete changes owing to constraints imposed by the hardware structure of the RIS.
Therefore, we focus on the codebook for 1-bit RIS whose elements have two possible configurations, denoted by a logic value of “-1” or “1”.  
Building upon the implicit representation of the 1-bit codebook, we formulate the problem as a multi-label classification task for the binarization of codebook elements.
Throughout the training process, the $6L$ dimensional vectors are employed as the MLP input, while the $M+N-1$ vectors, denoted as $p(\Phi)$, serve as the labels. The $\mathbf{q}$ signifies the length of $M+N-1$ vectors predicted by the INR model. The Binary Cross Entropy (BCE)~\cite{zhang2013review} loss is then utilized for back propagation, denoted as
\begin{equation*}
  \begin{aligned}
  \ell(p(\Phi),\mathbf{q})=-\sum_{i=1}^{M+N-1}\left[p_i\log(q_i)+(1-p_i)\log(1-q_i)\right],
  \label{equa:BCE loss}
  \end{aligned}
\end{equation*}
where $p_i$ denotes the $i$-th element of $p(\Phi)$, while $q_i$ represents the $i$-th element of $\mathbf{q}$.

\subsubsection{Labels Acquisition from Traversal Algorithm}
To ensure stable, rapid, and high-performance output labeling, we adopt an alternative method known as the Greedy Fast traversal algorithm~\cite{pei2021ris}. This algorithm is initialized with an all ``-1'' codebooks of size $M\times N$, sequentially altering the arrangement of all codebooks elements on each row and column from ``-1'' to ``1'' or ``1'' to ``-1'', and assessing the received signal strength following each alteration. If the modification enhances the received signal strength, the algorithm retains the process; otherwise, it reverts the change. This iterative process continues until the received signal strength stabilizes, signifying the algorithm has achieved the optimal codebooks. Extensive experimental findings demonstrate that the traversal algorithm delivers relatively swift processing and high-performance outcomes~\cite{pei2021ris}.

\subsubsection{Encoding and Decoding Scheme}
Directly utilizing the INR method to predict \(M \times N\) codebook is extremely challenging. Thus, it is crucial to reduce the number of neurons in the output layer. Following the principle of traversal algorithm, all the codebook elements are initially set to ``-1'' before the traversal commences. Denote this initial state as $\Phi_{0} \in \mathbb{C}^{M \times N}$. If the inverse operation is executed on the \(x\)-th $(1 \leq x \leq M)$ row, it can be recorded as $\Phi^{*}=D_{x}\times \Phi$, where $D_{x} \in \mathbb{C}^{M \times M}$ is a diagonal matrix with the \(x\)-th diagonal entry as  ``-1'' and all other as  ``1''. Similarly, the \(y\)-th $(1 \leq y \leq N)$  column is inverted, it denotes as $\Phi^{*}=\Phi \times D_{y}$, where $D_{y} \in \mathbb{C}^{N \times N}$ is a diagonal matrix with the \(y\)-th diagonal entry as ``-1" and all other elements as ``1". 

Since multiplication of diagonal matrices is commutative, it holds that
\begin{equation}
D_{x}^{2 n}=E, \quad
D_{x}^{2 n+1}=D_{x},
\end{equation}
where $n=0,1,2 \cdots $. Upon the completion of the entire traversal algorithm, the relationship between the codebook $\Phi_{opt}$ obtained and the initial codebook $\Phi_{0}$  is expressed as
\begin{equation}
\Phi_{opt}= \prod_{x}^{M} D_{x} ^{I+1} \times \Phi_{0} \times \prod_{y}^{N} D_{y}^{I+1},
\end{equation}
where $I$ is a binary indicator variable taking values of 0 or 1. With the help of $I$, the matrix 
$\prod_{x}^{M} D_{x} ^{I+1} $ represents whether the 1st, the 2nd, $\cdots$, the \(M\)-th rows of $\Phi_{0}$ are inverted, and the matrix $\prod_{y}^{N} D_{y} ^{I+1}$ signifies whether the 1st, the 2nd, $\cdots$ , the \(N\)-th columns of $\Phi_{0}$ are inverted. Therefore, it can be inferred that the $M\times N$ codebook can be represented using only $M+N$ variables.
 
\begin{figure}
  \centering
  \includegraphics[width = 0.45\textwidth, trim=170 80 160 80, clip]{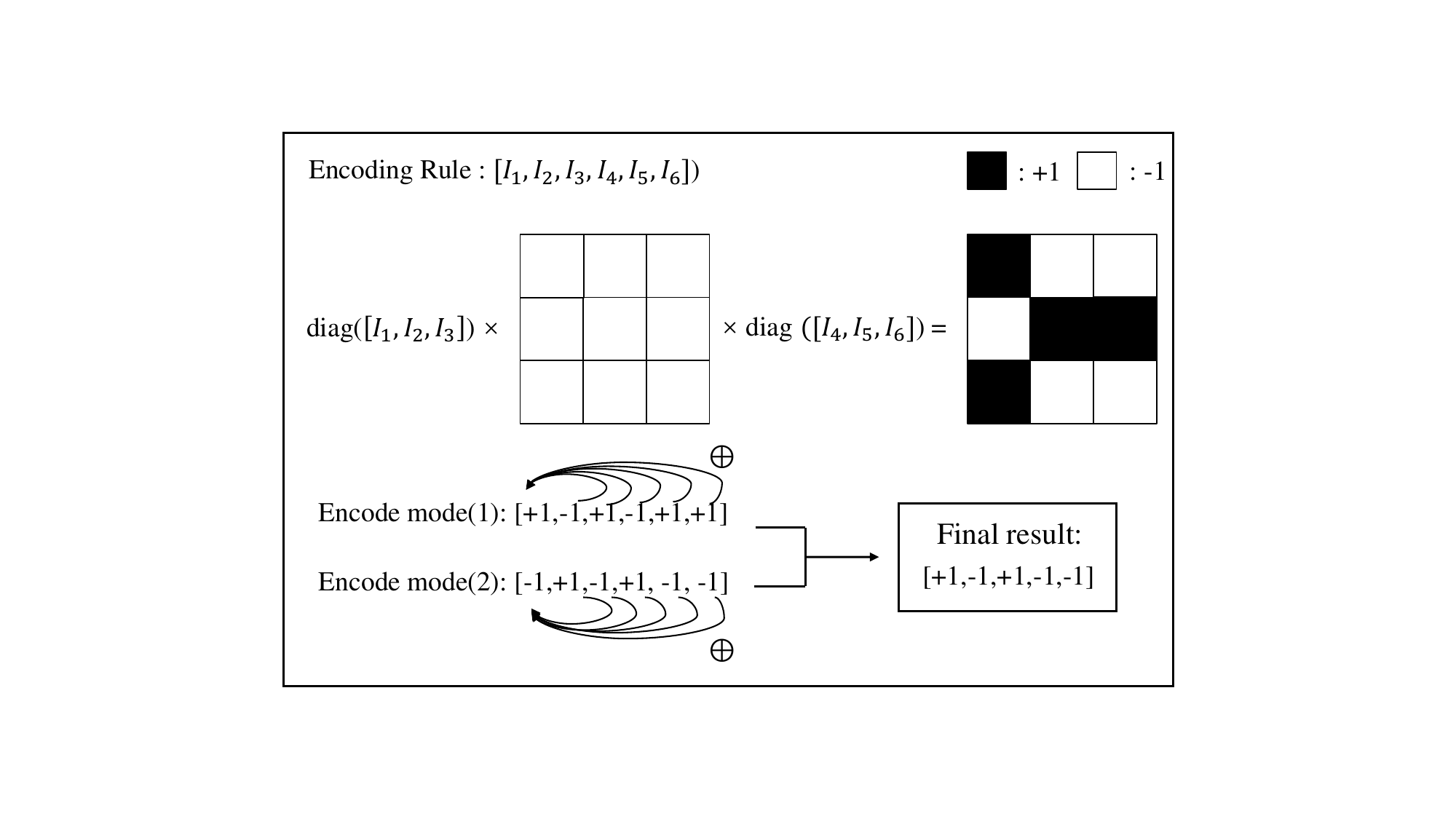}
  \caption{An illustration of the proposed encoding scheme: a $3 \times 3$ 1-bit RIS example. Specifically, we employ diagonal matrices $diag\left ( I_{1}, I_{2}, I_{3}\right )$ and $diag\left ( I_{4}, I_{5}, I_{6}\right )$ to represent row and column traversal, respectively. We find that both encoding mode(1): $[+1,-1,+1,-1,+1,+1]$ and mode(2): $[-1,+1,-1,+1,-1,-1]$ yield the same optimal codebooks. To further reduce the dimension, XOR operation is applied between the first bit of "+1" and the remaining 5 bits of $[-1,+1,-1,+1,+1]$, the same operation applies to the encode mode(2). In the end, encoding result of $[+1,-1,+1,-1,-1]$ can uniquely represent the optimal codebooks for $3 \times 3$ 1-bit RIS. That is, the predicted codebook dimension decreases from $M\times N$ to $M+N-1$ and finally to $M+N-1$.}
\label{fig:Encoding-Decoding scheme.pdf}
\vspace{-0.2cm}
\end{figure}

Furthermore, we conduct an XOR operation between the first bit of the codebook and the remaining $M+N-1$ codebooks bits, as illustrated in Fig.~\ref{fig:Encoding-Decoding scheme.pdf}. Ultimately, we can use only $M+N-1$ variables to represent the codebook of dimensions $M\times N$, and then decode results back to initial codebook through an inverse transformation. This encoding method is lossless and increases training efficiency.

\subsection{Model Training and Inference}
\begin{figure}
  \centering
  \includegraphics[width = 0.55\textwidth, trim=460 135 205 73, clip]{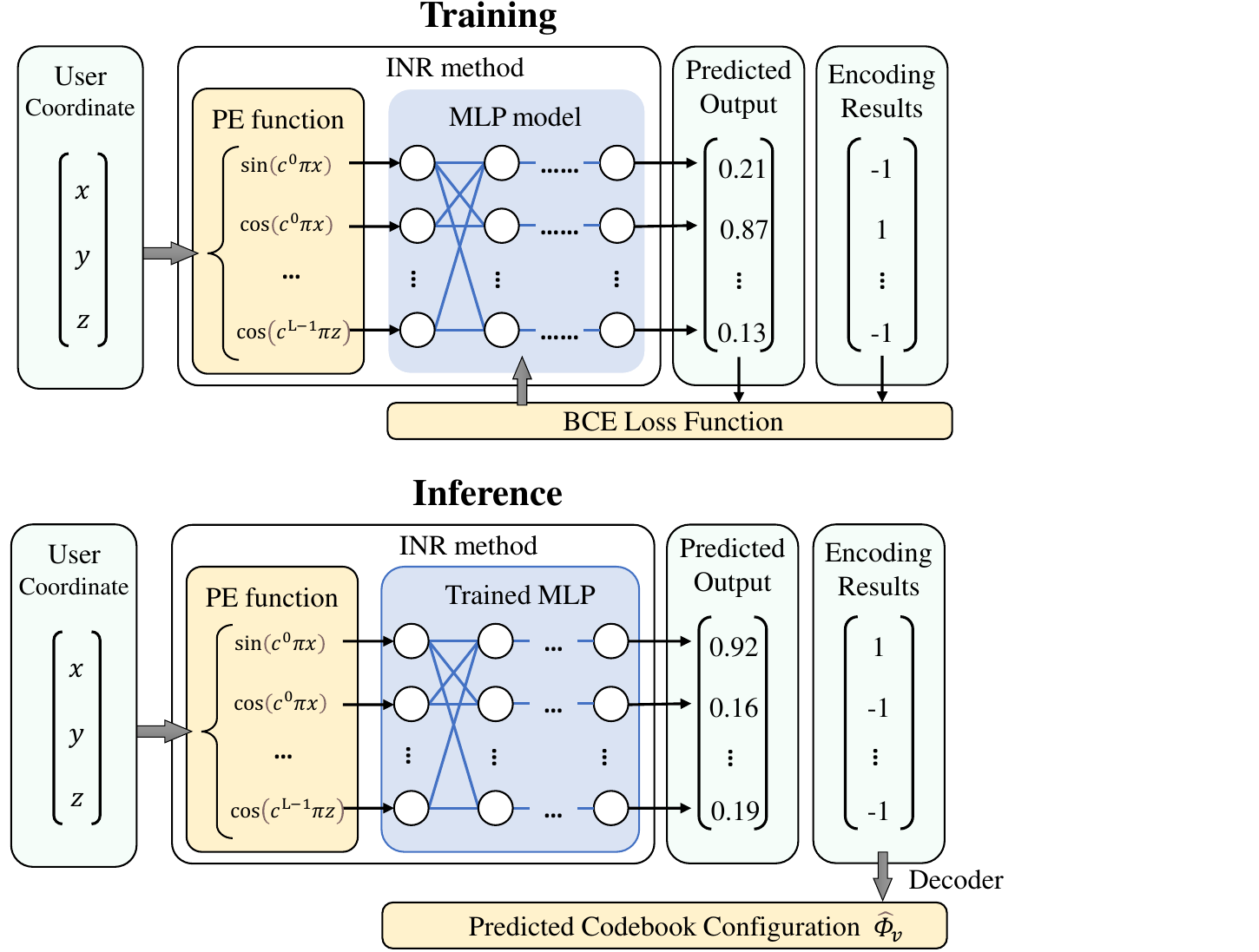}
  \caption{The framework includes two parts: the training step and the inference step. In the training step, INR model is trained on a dataset comprising user coordinates and the encoded codebook pairs. In the reference step, input solely the user coordinates to produce the optimal codebook.}
\label{fig:overall solution.pdf}
\vspace{-0.2cm}
\end{figure}

Our framework is depicted in Fig.~\ref{fig:overall solution.pdf}, which consists of two segments: the training step and the inference step.

\subsubsection{Training Step}
We symbolize the INR model as $f_{\theta}(\mathbf{v}) = \mathbf{q}$, where $\theta$ denotes the learnable parameters, $\mathbf{v}$ represents the user's coordinates and $\mathbf{q}$ is the $M+N-1$ vectors predicted by the INR model. The training step is formulated as follows:
\begin{align*}
\arg \min _{\theta} \quad \ell\left(f_{\theta}(\mathbf{v})-\operatorname{Enc}\left(\Phi_{\mathbf{v}}^{*}\right)\right).
\end{align*}
In this scenario, $\Phi_\mathbf{v}^*$ represents the ground-truth codebooks intended for the user coordinate $\mathbf{v}$, while $\operatorname{Enc}(\cdot)$ signifies the lossless encoding mechanism capable of converting the $M \times N$-size $\Phi_\mathbf{v}^*$ into a binary vector of length $M+N-1$. During the training step, our aim is to ensure that the output $f_{\theta}(\mathbf{v})$ of the INR model corresponds closely to the optimal codebooks encoding outcome $\operatorname{Enc}(\Phi_\mathbf{v}^*)$, achieved through the adjustment of the weights $\theta$. We utilize the Stochastic Gradient Descent (SGD) algorithm to update the parameters of the MLP.

\subsubsection{Inference Step}
Once we have acquired a trained INR model $f_\theta$, we articulate the inference step as follows:
\begin{equation*}
\mathbf{q}_{\text{real}} = f_{\theta}(\mathbf{v}_{\text{real}}), \quad
{\Phi}_{\mathbf{v}} = \text{decoder}(\epsilon(\mathbf{q}_{\text{real}})),
\end{equation*}
where $\mathbf{v}_{real}$ denotes the coordinate of the actual user, $\mathbf{q}_{real}$ represents the forecast output of the trained INR model, the function $\epsilon(\cdot)$ is defined such that it maps inputs within the range of 0 to 0.5 to ``-1'' and inputs within the range of 0.5 to 1 to ``1''. Additionally, $\operatorname{decoder}(\cdot)$ stands for the corresponding decoding mechanism, capable of restoring $M \times N$ dimensional codebooks to $\hat{\Phi}\mathbf{v}$ .
\begin{figure*}
  \centering
  \includegraphics[width = 0.9\textwidth, trim=20 150 15 140, clip]{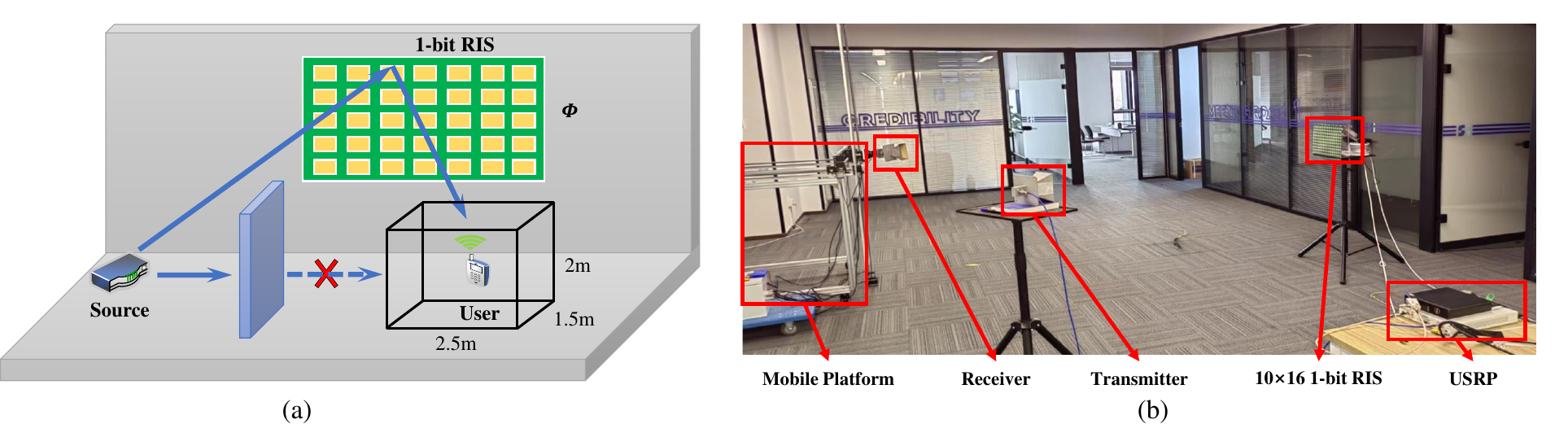}
  \caption{(a) The simulation scenario. (b) The experimental scenario.}
\label{fig:Simulated and measured scenarios.pdf}
\vspace{-0.2cm}
\end{figure*}

\section{Performance Evaluation}
In this section, experiments on both simulated and measured data are constructed to evaluate the effectiveness of the proposed INR method. Important parameter setting are outlined in Table~\ref{tab:parameters}. In particular, we establish both simulated and experimental environment, constructing the optimal codebook dataset through the traversal algorithm. For comparison, we utilize four benchmarks: 
\begin{itemize}
    \item \textbf{DNN }: DNN method\cite{huang2019indoor}.
    \item \textbf{APX}: Approximation algorithm\cite{zhang2022configuringAPX}.
    \item \textbf{SDR}: Semi-definite relaxation\cite{wu2019intelligentSDR}.
    \item \textbf{Manopt}: Manifold optimization\cite{xiong2023risMAN}.
\end{itemize}

\subsection{Dataset Construction}\label{subsec:Dataset Construction}

To compile the training dataset, we establish $K$ sampling points $\mathbf{v}_1, \mathbf{v}_2, ..., \mathbf{v}_K$ uniformly across the space where user may be present. We then leverage the traversal algorithm to derive the optimal codebooks as the label and assemble the dataset $D:=\left\{(\mathbf{v}_1, \Phi_\mathbf{v_1}^*), (\mathbf{v}_2, \Phi_\mathbf{v_2}^*), ..., (\mathbf{v}_K, \Phi_\mathbf{v_K}^*)\right\}$.

\subsubsection{Simulation and Measured Dataset}
We create a simulated setting, as illustrated in Fig. \ref{fig:Simulated and measured scenarios.pdf}(a), 
where the user can only receive the signal through the $10\times10$ 1-bit RIS installed on the wall.
The user exists within a three-dimensional space measuring $2.5m \times 1.5m \times 2m$ and we sample a total of  $20\times15\times15=4500$ points. Draw upon the aspect of electromagnetic field presented in~\cite{yang20161}, we model a RIS-aided system that can compute the signal strength at each user's position. By utilizing the proposed traversal algorithm, we are able to derive the optimal codebook corresponding to any point, facilitating the construction of the simulation dataset. The experimental scenario is depicted in Fig. \ref{fig:Simulated and measured scenarios.pdf}(b). The setup incorporates a $10\times16$ 1-bit RIS and a USRP device, enabling the generation of signals. Additionally, we use two horn antennas capable of transmitting or receiving signals in appointed directions. Moreover, the mobile platform offers an accessible space measuring $0.9m \times 0.6m \times 1.4m$, establishing a grid of $10\times7\times15=1050$ points. Similarly, The measured dataset is constructed based on the traversal algorithm.

\subsubsection{Dataset Processing} Before commencing the training of the MLP, the dataset $D$ goes through a preprocessing phase. The first step involves scaling the coordinate data to fit within the interval $[-1, +1]$. Subsequently, the scaled data is transformed into a $6L$ dimensional vector as the input of MLP, accomplished by applying the PE technique $\Gamma(\cdot)$.  Moreover, the encoding scheme $p(\cdot)$ is employed to encode the codebooks within $D$. This results in the generation of binary vectors, with a length of $M+N-1$, serving as the labels for the MLP training process.

\subsection{Evaluation Metrics}

The evaluation criteria mainly consist of two parts: The first part involves the user's received signal strength, with a stronger signal indicating a better codebook configuration. Under the condition of maintaining other factors consistent, This part involves comparing the signal strengths obtained through the proposed method with four other benchmark methods.
The second aspect pertains to the average accuracy of the model's predicted codebooks, referring to the precision of the codebook elements predicted by the INR model method. The accuracy evaluation is defined as:
\begin{equation*}
  \text{Accuracy} = \frac{\text{correct predictions of codebooks }}{\text{total predictions of codebooks}  }\times 100\%
  \label{equa:accuracy}
\end{equation*}




\begin{table}
  \centering
  \caption{Key Network Parameters}
    \begin{tabular}{l|l}
    \hline
    \textbf{Parameter} & \textbf{Value} \\
    \hline
    Base frequency  & $C=1.35 $ \\
    Length of PE & $L=40 $ \\
    Dimension of input   & $6\times L$ \\
    Number of hidden layer & $L_h=4$ \\
    width of hidden layer  & $W_h=128$ \\
    optimizer & SGD \\
    Batchsize & $\text{Batchsize}=64$ \\
    Learning rate & $10^{-3}$ \\
    \hline
    \end{tabular}%
  \label{tab:parameters}%
\end{table}%

\subsection{Performance Evaluation}
\subsubsection{Comparison with Benchmarks}
The measurements of received signal strength for each method are detailed in Table~\ref{tab:Comparison experiments}. Both the simulated and measured data were used for experimentation across all methods. Notably, learning-based approach exhibited superior performance compared to the model-based approach in general. Furthermore, the INR method surpassed all other benchmarks across diverse datasets.
In contrast to the model-based methods, the INR method showcased an average performance improvement of around $7 \text{ dBm}$ in terms of received power. This significant performance gain highlights the efficacy of the INR-based approach, emphasizing its capacity to enhance signal strength.

\begin{table}
  \centering
  \caption{Received signal strength from various methods}
  \resizebox{0.45\textwidth}{!}{
    \begin{tabular}{cccc}
    \toprule
    
        & Methods & \makecell[c]{Simulation result \\ (dBm)}   &\makecell[c]{Measured result \\ (dBm)}  \\
        \midrule
            
            \multirow{3}[0]{*}{Model-based} 
                & APX \cite{zhang2022configuringAPX}  & -23.93  & -29.40  \\
                & SDR \cite{wu2019intelligentSDR}  & -23.89  & -29.24  \\
                & Manopt \cite{xiong2023risMAN}  & -23.80 & -29.27 
                \\
                \midrule
            \multirow{2}[0]{*}{Learning-based}
                & DNN \cite{huang2019indoor} & -20.28  & -28.85  \\
                & \textbf{Ours}  & \textbf{-16.85} & \textbf{-26.43} \\
    \bottomrule
    \end{tabular}%
    }
    \label{tab:Comparison experiments}%
\end{table}

\begin{figure}
  \centering
  \includegraphics[width = 0.40\textwidth, trim=20 220 40 180, clip]{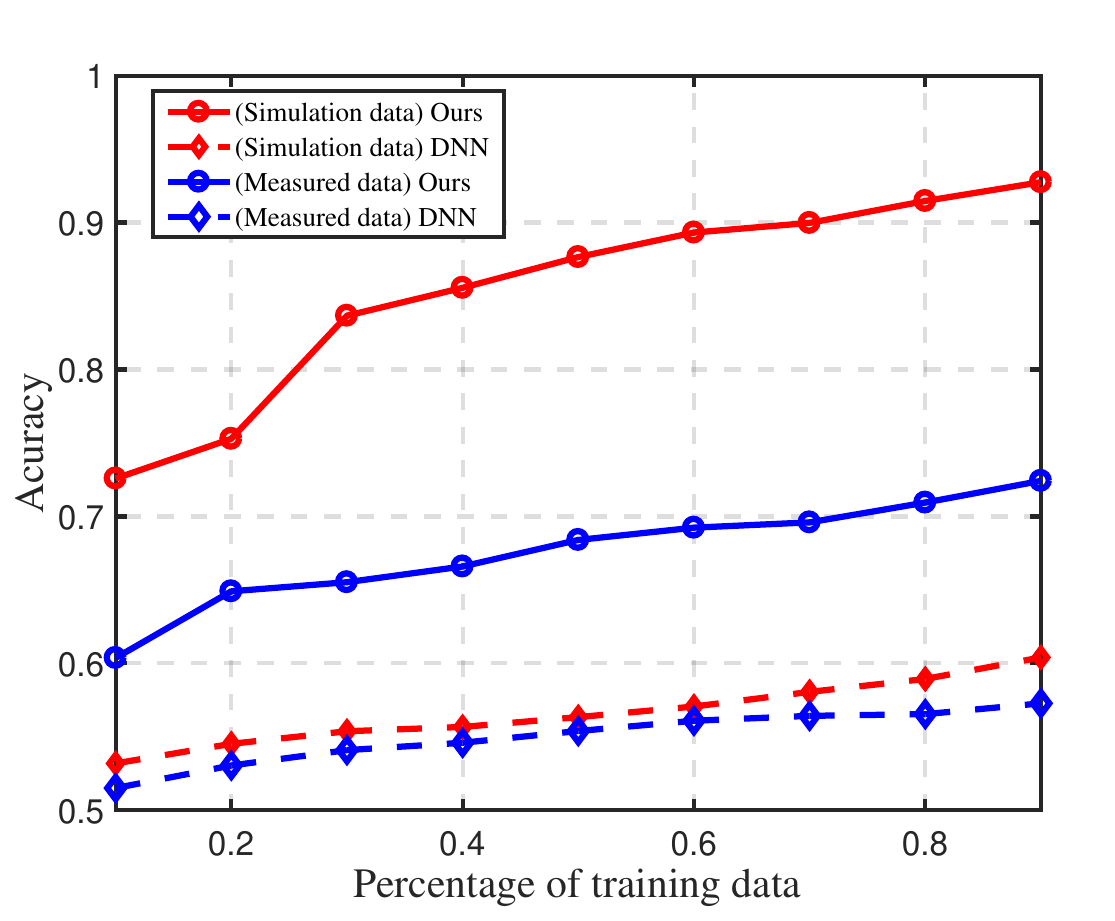}
  \caption{The accuracy  w.r.t the percentage of training data.}
\label{fig:plt.eps}
\vspace{-0.4cm}
\end{figure}

Furthermore, we conducted experiments to analyze the performance development of various methods concerning the percentage of training data. Fig.~\ref{fig:plt.eps} presents a visual comparison of the accuracy results between our proposed method and the DNN. Notably, in the simulated dataset, INR-based algorithm achieved an impressive accuracy of 83.7\% using only 30\% of the available training data. This outperformance was notable in contrast to the DNN, which achieved an accuracy of 65.5\%.
This considerable discrepancy in performance can be attributed to the advanced capability of the position encoding method in grasping intricate information and the encoding strategy to enhance learning efficiency.

\subsubsection{Ablation Experiments}
To evaluate the individual contributions of specific components within our approach, we conducted ablation experiments. The results, showcased in Table~\ref{tab:ablation study}, illustrate that the integration of ReLU activation, positional and codebook encoding surpasses alternative methods. Notably, the inclusion of PE enriches the network's expressiveness by enhancing the dimensionality of the input coordinates. Concurrently, the utilization of codebook encoding technique reduces the dimension of the output space, enhancing the learning efficiency. These combined effects notably bolster the overall performance of the neural network. Furthermore, our comparative analysis of various activation functions has highlighted the superiority of ReLU in terms of accuracy.

\section{Conclusion}\label{sec:Conclusion}


We investigate the configuration of codebooks for RIS-aided system. 
From the perspective of signal radiation mechanisms, we define a implicit relationship between the user's coordinates and the codebook, and propose a novel INR method to solve this implicit mapping problem. 
It is the first time that the INR technique has been introduced to the task of the codebook configuration of RIS.
Additionally, considering the significant practical applications of 1-bit RIS, we formulate the 1-bit codebook configuration as a multi-label classification problem, and propose an encoding-decoding strategy to minimize codebook dimensions, thereby enhancing the training efficiency. Empirical evidence from simulated and measured data demonstrates the substantial advantages of our approach. 
Further extending the proposed method to scenarios with multiple users is a promising future direction.

\begin{table}
  \centering
  \caption{Ablation experiments}
    \begin{tabular}{cccccc}
    \toprule
    \multirow{2}[2]{*}[0pt]{Activation} & \multirow{2}[2]{*}[0pt]{PE} & \multicolumn{2}{c}{Simulation data} & \multicolumn{2}{c}{Measured Data} \\
    \cmidrule(r){3-4} \cmidrule(r){5-6}
         &       & enc   & w/o enc & enc   & w/o enc \\
    \midrule
    \multirow{2}[1]{*}{ReLU} & \ding{55} & 74.33\% & 58.37\% & 65.43\% & 56.42\% \\
                             & \ding{51} & \textbf{91.46\%} & 87.06\% & \textbf{70.94\%} & 64.41\% \\
    \midrule
    \multirow{2}[0]{*}{LeakyReLU} & \ding{55} &74.26\% & 58.30\% & 67.56\% & 59.13\% \\
                            & \ding{51} & 90.77\% & 87.58\% & 68.97\% & 63.64\% \\
    \midrule
    \multirow{2}[1]{*}{GELU} & \ding{55} & 74.40\% & 61.38\% & 65.98\% & 56.45\% \\
                             & \ding{51} & 90.88\% & 87.36\% & 69.94\% & 60.84\% \\
    \midrule
    \multirow{2}[0]{*}{Sin} & \ding{55} & 72.06\% & 54.66\% & 56.37\% & 52.84\% \\
                             & \ding{51} & 88.81\% & 83.62\% & 57.24\% & 53.19\% \\   
    \midrule
    \multirow{2}[0]{*}{Tanh} & \ding{55} & 74.33\% & 57.60\% & 64.75\% & 59.91\% \\
                                  & \ding{51} & 89.70\% & 86.85\% & 68.62\% & 61.79\% \\
       \bottomrule
    \end{tabular}%
  \label{tab:ablation study}%
\end{table}%
\vspace{-5pt}


\bibliographystyle{IEEEtran}
\bibliography{reference}

\begin{thebibliography}{10}
\providecommand{\url}[1]{#1}
\csname url@samestyle\endcsname
\providecommand{\newblock}{\relax}
\providecommand{\bibinfo}[2]{#2}
\providecommand{\BIBentrySTDinterwordspacing}{\spaceskip=0pt\relax}
\providecommand{\BIBentryALTinterwordstretchfactor}{4}
\providecommand{\BIBentryALTinterwordspacing}{\spaceskip=\fontdimen2\font plus
\BIBentryALTinterwordstretchfactor\fontdimen3\font minus \fontdimen4\font\relax}
\providecommand{\BIBforeignlanguage}[2]{{%
\expandafter\ifx\csname l@#1\endcsname\relax
\typeout{** WARNING: IEEEtran.bst: No hyphenation pattern has been}%
\typeout{** loaded for the language `#1'. Using the pattern for}%
\typeout{** the default language instead.}%
\else
\language=\csname l@#1\endcsname
\fi
#2}}
\providecommand{\BIBdecl}{\relax}
\BIBdecl

\bibitem{wu2019towards}
Q.~Wu and R.~Zhang, ``Towards smart and reconfigurable environment: Intelligent reflecting surface aided wireless network,'' \emph{IEEE Communications Magazine}, vol.~58, no.~1, pp. 106--112, 2019.

\bibitem{di2020smart}
M.~Di~Renzo, A.~Zappone \emph{et~al.}, ``Smart radio environments empowered by reconfigurable intelligent surfaces: How it works, state of research, and the road ahead,'' \emph{IEEE journal on selected areas in communications}, vol.~38, no.~11, pp. 2450--2525, 2020.

\bibitem{zhang2022configuringAPX}
Y.~Zhang, K.~Shen \emph{et~al.}, ``Configuring intelligent reflecting surface with performance guarantees: Optimal beamforming,'' \emph{IEEE Journal of Selected Topics in Signal Processing}, vol.~16, no.~5, pp. 967--979, 2022.

\bibitem{wu2019intelligentSDR}
Q.~Wu and R.~Zhang, ``Intelligent reflecting surface enhanced wireless network via joint active and passive beamforming,'' \emph{IEEE Transactions on Wireless Communications}, vol.~18, no.~11, pp. 5394--5409, 2019.

\bibitem{xiong2023risMAN}
R.~Xiong, J.~Zhang \emph{et~al.}, ``{RIS}-aided wireless communication in real-world: Antennas design, prototyping, beam reshape and field trials,'' \emph{arXiv preprint arXiv:2303.03287}, 2023.

\bibitem{abdallah2023deep}
A.~Abdallah, A.~Celik \emph{et~al.}, ``Deep reinforcement learning based beamforming codebook design for ris-aided mmwave systems,'' in \emph{2023 IEEE 20th Consumer Communications \& Networking Conference (CCNC)}.\hskip 1em plus 0.5em minus 0.4em\relax IEEE, 2023, pp. 1020--1026.

\bibitem{huang2019indoor}
C.~Huang, G.~C. Alexandropoulos \emph{et~al.}, ``Indoor signal focusing with deep learning designed reconfigurable intelligent surfaces,'' in \emph{2019 IEEE 20th international workshop on signal processing advances in wireless communications (SPAWC)}.\hskip 1em plus 0.5em minus 0.4em\relax IEEE, 2019, pp. 1--5.

\bibitem{sitzmann2020implicit}
V.~Sitzmann, J.~Martel \emph{et~al.}, ``Implicit neural representations with periodic activation functions,'' \emph{Advances in Neural Information Processing Systems}, vol.~33, pp. 7462--7473, 2020.

\bibitem{FourierMapping}
M.~Tancik, P.~Srinivasan \emph{et~al.}, ``Fourier features let networks learn high frequency functions in low dimensional domains,'' \emph{Advances in Neural Information Processing Systems}, vol.~33, pp. 7537--7547, 2020.

\bibitem{mi2023towards}
T.~Mi, J.~Zhang \emph{et~al.}, ``Towards analytical electromagnetic models for reconfigurable intelligent surfaces,'' \emph{IEEE Transactions on Wireless Communications}, 2023.

\bibitem{tucker2020single}
R.~Tucker and N.~Snavely, ``Single-view view synthesis with multiplane images,'' in \emph{Proceedings of the IEEE/CVF Conference on Computer Vision and Pattern Recognition}, 2020, pp. 551--560.

\bibitem{mildenhall2021nerf}
B.~Mildenhall, P.~P. Srinivasan \emph{et~al.}, ``Nerf: Representing scenes as neural radiance fields for view synthesis,'' \emph{Communications of the ACM}, vol.~65, no.~1, pp. 99--106, 2021.

\bibitem{orekondy2022winert}
T.~Orekondy, P.~Kumar \emph{et~al.}, ``Winert: Towards neural ray tracing for wireless channel modelling and differentiable simulations,'' in \emph{The Eleventh International Conference on Learning Representations}, 2022.

\bibitem{xiong2022optimal}
R.~Xiong, X.~Dong \emph{et~al.}, ``Optimal discrete beamforming of reconfigurable intelligent surface,'' \emph{arXiv preprint arXiv:2211.04167}, 2022.

\bibitem{pei2021ris}
X.~Pei, H.~Yin \emph{et~al.}, ``Ris-aided wireless communications: Prototyping, adaptive beamforming, and indoor/outdoor field trials,'' \emph{IEEE Transactions on Communications}, vol.~69, no.~12, pp. 8627--8640, 2021.

\bibitem{zhang2013review}
M.-L. Zhang and Z.-H. Zhou, ``A review on multi-label learning algorithms,'' \emph{IEEE Transactions on Knowledge and Data Engineering}, vol.~26, no.~8, pp. 1819--1837, 2013.

\bibitem{yang20161}
H.~Yang, F.~Yang \emph{et~al.}, ``A 1-bit $10\times 10$ reconfigurable reflectarray antenna: design, optimization, and experiment,'' \emph{IEEE Transactions on Antennas and Propagation}, vol.~64, no.~6, pp. 2246--2254, 2016.

\end{thebibliography}

\end{document}